\documentclass[12pt,a4wide]{article}

\usepackage{epsfig,lscape}
\usepackage{fancyhdr}	
\usepackage{amsmath, amsfonts}

\usepackage{natbib}

\usepackage{times}

\usepackage{enumerate}
\newtheorem{defn}{Definition}[section]

\newtheorem{thm}[defn]{Theorem}

\newtheorem{assu}[defn]{Assumption}
\newtheorem{example}[defn]{Example}

\newcommand{\cip}{\mbox{$\perp\!\!\!\perp$}}
\def\T{{ \mathrm{\scriptscriptstyle T} }}
\begin{document}

\title{Defining and estimating causal direct and indirect effects when setting the mediator to specific values is not feasible}




\author{JUDITH J.~LOK$^\ast$\\
Department of Biostatistics,\\
Harvard T.H.~Chan School of Public Health,\\
655 Huntington Avenue, Boston, MA 02115, USA;\\
{jlok@hsph.harvard.edu}}

\markboth%
{Judith J. Lok}
{Organic direct and indirect effects}

\maketitle


\begin{abstract}{Natural direct and indirect effects decompose the effect of a treatment into the part that is mediated by a covariate (the mediator) and the part that is not. Their definitions rely on the concept of outcomes under treatment with the mediator ``set'' to its value without treatment. Typically, the mechanism through which the mediator is set to this value is left unspecified, and in many applications it may be challenging to fix the mediator to particular values for each unit or individual. Moreover, how one sets the mediator may affect the distribution of the outcome. This article introduces ``organic'' direct and indirect effects, which can be defined and estimated without relying on setting the mediator to specific values. Organic direct and indirect effects can be applied for example to estimate how much of the effect of some treatments for HIV/AIDS on mother-to-child transmission of HIV-infection is mediated by the effect of the treatment on the HIV viral load in the blood of the mother.}
\end{abstract}




\noindent Causal inference, Direct and indirect effect, HIV/AIDS,
  Mediation, Observational study, Organic direct and indirect effect.



\section{Introduction}
\label{Intro}

Researchers are often interested in investigating the mechanisms
behind effective treatments or exposures. The topic of which part of
the effect of a treatment is ``mediated'' by a covariate is of
particular importance. The mediated part of a treatment effect is due
to treatment induced changes in a ``mediator'' covariate $M$. This is
the so-called indirect effect; as opposed to the so-called direct
effect, not mediated by covariate $M$. Mediation analysis has gained
much prominence in methodological and empirical research in recent
years. Mediation analysis is particularly popular in the health sciences,
like epidemiology and psychology. In June 2015, \cite{BaronKenny} has over 52.000 citations in Google Scholar, many of them after 2010. Therefore, clarifying the
assumptions required for mediation analysis is paramount.

A recent literature has provided a rigorous theoretical framework for
the definition and estimation of causal direct and indirect effects, see e.g.\ \cite{RobGreenmed},
\cite{Pearlmed,Pearlmed2}, \cite{Imai}, \cite{Tyler},
\cite{JamieThomas}, \cite{Ericmedsurv}. In this
literature, the controlled direct effect is the effect of a treatment
had the mediator been set to a pre-determined value. The natural
direct effect is the effect of a treatment had the mediator been set
to the value that it would have taken without treatment. Thus, the
natural direct effect for a particular unit or patient can be
represented by $Y_{1,M_0}-Y_0$: the difference between the unit's
outcome $Y_{1,M_0}$ under treatment had the mediator been set to the
value $M_0$ it would have taken without treatment and the unit's
outcome $Y_0$ without treatment. Notice that the mediator is kept
constant at $M_0$, so this is the natural direct effect not mediated
by $M$. Similarly, the natural indirect effect can be represented by
$Y_1-Y_{1,M_0}$, where $Y_1$ is the outcome under treatment.
 It has been argued that to study the causal mechanisms by which
 particular treatments are effective, natural direct and indirect
 effects are more relevant than controlled direct effects
 (\cite{Pearlmed}, \cite{Tyler}). \cite{Pearlmed} also notes that
 controlled indirect effects are not defined.

As seen from the definition of natural direct and indirect effects,
one needs ``cross-worlds'' quantities in order to define natural direct and indirect effects.
In particular, $Y_{1,M_0}$ is the outcome under
treatment but with the mediator set to the value, $M_0$, it would have
taken without treatment. In practice, it may be rare that mediators
take the same value with treatment, $M_1$, as without treatment,
$M_0$. Even if this happened for specific sample units (e.g.,
patients; from now on: units), it would be impossible to identify
those units. Under treatment, the value of the mediator without
treatment is not observed, so it is unclear to which value the
mediator should be set for any particular unit. In addition,
identification of natural direct and indirect effects relies on
assumptions about the outcomes $Y_{a,m}$ of a unit under all
combinations of the treatment and the mediator
(\cite{Pearlmed,Pearlmed2}, \cite{Tyler}, \cite{Imai},
\cite{Ericmedsurv}). This is a problem in many practical applications:
setting the mediator to particular values is often not feasible if the
mediator is not a treatment itself. If setting the mediator to a
particular value $m$ is not feasible, the interpretation of the
$Y_{a,m}$ is unclear. However, the common approaches to causal
mediation analysis all rely on the existence of all
$Y_{a,m}$ (\cite{JamieThomas}).

To overcome these problems, this article proposes instead to base
mediation analysis on newly defined ``organic'' interventions ($I$) on
the mediator. Organic interventions $I$ cause the mediator to have a
specific distribution: the distribution of the mediator without
treatment, given pre-treatment common causes of mediator and
outcome. An organic intervention could be an additional treatment that affects the distribution of the mediator. Theorem~\ref{gen} shows that organic direct and indirect effects are often generalizations of natural direct and indirect effects and the direct and indirect effects introduced in \cite{Vanessa2}. Like the current article, \cite{Geneletti} considers interventions on the mediator, but does not condition on common causes $C$ of mediator and outcome, unless the interest is in effects conditional on $C$ or in effects when $C$ is manipulated. \cite{RobGreenmed}, \cite{Pearlmed}, and, for organic interventions, Section~\ref{definition} argue that ignoring $C$ can produce invalid estimators. Also Section~\ref{unique} on uniqueness of organic direct and indirect effects requires that $C$ is taken into account.

Most of the causal inference literature on mediation has adopted the so-called cross-worlds assumption, an assumption involving the joint distribution of counterfactuals under different values of the treatment (\cite{Pearlmed,Pearlmed2}, \cite{Tyler}, \cite{Imai}, \cite{Ericmedsurv}). An issue with this assumption is that it can never be tested or imposed by design, not even in a clinical trial where the treatment and the mediator can both be set to any desired value by the experimenter. In contrast, whether "setting the mediator" is an organic intervention can be tested in such a clinical trial.

The theory in this article turns out to lead to the same
numerical results as introduced by other authors,
\cite{BaronKenny} and, more recently, e.g. \cite{Pearlmed,Pearlmed2},
\cite{Tyler}, \cite{Imai}, \cite{EricTylerMedAnnals}, 
and \cite{Vanessa2}. This article thus provides an interpretation of existing numerical results when the mediator cannot be set.


Proofs can be found in Web-appendix~A. I illustrate the usefulness of organic direct and indirect effects as
opposed to natural direct and indirect effects in two examples: 1.\ the smoking and low birth weight paradox, see Web-appendix~B and 2.\ the effect of AZT, a drug used for the treatment of HIV, on
mother-to-child transmission of HIV-infection. I investigate how much
of the effect of AZT is mediated by the HIV viral load, the amount of
HIV-virus in the blood of the mother.


\section{Setting and notation}
\label{setting}

For ease of exposition, I first consider randomized treatments. Section~\ref{observational} extends the analysis to
non-randomized treatments. For each unit, observables include the
following quantities. $A$ is the randomized treatment, which is $1$
for the treated and $0$ for the untreated. $C$ are pre-treatment
common causes of the mediator and the outcome. As noted by
\cite{RobGreenmed}, \cite{Pearlmed}, and others, these variables $C$
have to be taken into account in order to identify the natural direct
and indirect effect, even if the treatment is randomized. As will
become clear later, pre-treatment common causes $C$ also have to be
taken into account to identify the organic direct and indirect
effect. Like most of the literature on mediation, this article assumes
that there are no post-treatment common causes of the mediator and the
outcome. $M$ is the observed value of the mediator.  $Y$ is the
observed outcome. $Y_0$ is the (counterfactual) outcome without
treatment, and $Y_1$ is the (counterfactual) outcome under
treatment. Obviously, for each unit either $Y_1$ or $Y_0$ is observed,
but not both. Similarly, $M_0$ is the mediator without treatment, and
$M_1$ is the mediator under treatment. I assume that $C$ is observed
first, then $A$, then $M$, and then $Y$. The Directed Acyclic Graph
(DAG) of Figure~1 describes the set-up. In all of this article, it is assumed that observations and counterfactuals for the different units are independent and identically distributed.




\section{Natural direct and indirect effects: an overview}\label{natural}

Natural direct and indirect effects, introduced in \cite{RobGreenmed}
and \cite{Pearlmed}, are based on the outcome $Y_{1,M_0}$ under
treatment with the mediator set to the value it would have taken
without treatment, $M_0$.
The natural direct effect is defined as $Y_{1,M_0}-Y_0$. The natural
direct effect is not affected by changes in the value of the mediator
induced by the treatment, because for both $Y_{1,M_0}$ and $Y_0$ the
mediator is equal to $M_0$. The natural indirect effect is defined as
$Y_1-Y_{1,M_0}$. This is the mediated part: the only difference
between these quantities is the change in the value of the mediator,
$M_1$ versus $M_0$. 

To identify natural direct and indirect effects, previous authors
(e.g.\ \cite{Pearlmed,Pearlmed2}, \cite{RobGreenmed}, \cite{Tyler},
\cite{Imai} and \cite{Ericmedsurv}) assume the existence of
counterfactual outcomes under all possible combinations of the
treatment and the mediator, $Y_{a,m}$. In the words of
\cite{JamieThomas}, there has to be ``reasonable agreement'' as to
what is the ``closest possible world'' in which the mediator has a
specific value, a value which is different from the one that was
observed.  There are cases where reasonable agreement may exist. For
example, \cite{Pearlmed} describes a setting where the mediator is a
treatment, aspirin, in which case the mediator could be set to
specific values.  However, in many practical situations the mediator
of interest is not a treatment, and there is no known way in which one
can set the mediator to a specific value. Then, the
quantities $Y_{1,M_0}$ and $Y_{a,m}$ are not clearly defined. \cite{ColeF} provide a nice example: "there are many competing ways to assign (hypothetically) a body mass index of $25$ kg/m$^2$ to an individual, and each of them may have a different causal effect on the outcome".

The cross-worlds or mediator-randomization assumption that is
generally used to identify natural direct and indirect effects states
that
\begin{equation}\;\;\;\;\;\;\;\;\;Y_{a',m}\cip M_a \mid C=c, A=a.\label{crosseqn}\end{equation}
In words: for a unit with treatment $A=a$ and pre-treatment covariates
$C$, the mediator under treatment $a$ ($M_a$), should be independent
of the outcome under any other treatment-mediator combination
($Y_{a',m}$, the outcome under treatment $a'$ had the mediator been
set to $m$).
Identification of natural direct and indirect effects thus involves assumptions about
cross-worlds quantities. Suppose for now that the $Y_{a'm}$ are all
well-defined: there is reasonable agreement about how to set the
mediator to a specific value. Then (\ref{crosseqn}) is similar to the
classical assumption of no unmeasured confounding in causal inference
(e.g., \cite{Aids}). To understand (\ref{crosseqn}),
notice that ``nature'' determines the values of the mediators $M_a$
and the outcomes $Y_{a'm}$, based on $C$ and possibly other
factors. Equation~(\ref{crosseqn}) thus states that given $C$, the
$Y_{a'm}$ do not help to predict $M_a$; or, nature did not have more
information on the potential outcomes $Y_{a'm}$ to determine the value
of the mediator $M_a$ than recorded in $C$. In other words, all common
causes of mediator and outcome have to be recorded in $C$. Then, under a consistency assumption,
\begin{equation}\label{medf}E\left(Y_{1,M_0}\right)=\int_{(c,m)}E\left[Y\mid M=m,C=c,A=1\right]f_{M\mid C=c,A=0}(m)f_C(c)dm\,dc;
\end{equation}
the ``mediation formula'', see e.g.~\cite{Pearlmed, Pearlmed2}, \cite{Tyler}, and \cite{Imai}.

Under certain conditions (strong parametric assumptions, linear models
and no exposure-mediator interaction), the estimators for the natural
direct and indirect effects resulting from (\ref{medf}) are the same as
the estimators in \cite{BaronKenny}, the founding article on direct
and indirect effects. The causal inference literature on natural direct and indirect
effects thus generalizes the approach of
\cite{BaronKenny} and adds a causal interpretation to their
estimators.

\section{Definitions of organic intervention and organic direct and indirect effects}
\label{definition}

This section defines organic direct and indirect effects.
Analogously to natural direct and indirect effects, this article
focuses on interventions $I$ that cause the mediator under treatment $A=1$ and intervention $I$,
$M_{1,I=1}$, to have the same distribution as $M_0$, given the
pre-treatment common causes $C$ of mediator and outcome. However, for
individual units, $M_{1,I=1}$ does not need to be exactly the value the
mediator would have had without treatment, $M_0$. This is a
considerable relaxation, especially because this distribution can be
estimated from the observed data (provided $C$ has been measured), while
individual values of $M_0$ are not observed under treatment. Hence, it
is possible to imagine an intervention that leads to this
distribution. I term this type of interventions organic because they
depend on the entire distribution of $M_0$, the mediator without
treatment, rather than on individual values of $M_0$. Write $Y_{1,I=1}$ for the
outcome under treatment $A=1$ and intervention $I$. Then,
\begin{defn}\emph{(Organic intervention).}\label{mim}
An intervention $I$ is an organic intervention with respect to $C$
if
\begin{equation}\label{defint}
M_{1,I=1}\mid C=c \sim M_0 \mid C=c
\end{equation}
\begin{equation}\label{ident}
\;Y_{1,I=1}\mid M_{1,I=1}=m,C=c \sim Y_1\mid M_1=m,C=c,
\end{equation}
both hold, where $\sim$ indicates having the same distribution.
\end{defn}
Equation~(\ref{defint}) says that $I$ ``holds the mediator at its
distribution under no treatment'': given $C$, there is no difference in the
distribution of the mediator under treatment $A=1$ combined with intervention $I$ and
the distribution of the mediator under no
treatment. Rather than the cross-worlds assumption of
equation~(\ref{crosseqn}), I assume
equation~(\ref{ident}). Equation~(\ref{ident}) intuitively states that $I$ ``has no direct effect
on the outcome'': for units
with pre-treatment common causes of mediator and outcome fixed at
$C=c$, the prognosis of units under treatment ``with mediator $M_{1,I=1}$
being equal to $m$ under intervention $I$'' is the same as the
prognosis of units under treatment ``with mediator $M_1$ being equal
to $m$ without $I$''. In other words, given $C$,
treated units with mediator equal to $m$ ($M_1=m$) without intervention $I$ are
representative of treated units with $M_{1,I=1}=m$ under intervention $I$.
Equation~(\ref{ident}) could be relaxed by assuming instead that
$E\left[Y_{1,I=1}\mid M_{1,I=1}=m,C=c\right] = E\left[Y_1\mid
  M_1=m,C=c\right]$. If the intervention $I$ on the mediator has a
direct effect on the outcome, equation~(\ref{ident}) fails to
hold. Equation~(\ref{ident}) is related to the assumption of ``partial
exchangeability'' in \cite{RobGreenmed} and can be discussed with
subject matter experts (Web-appendix~E may also help).
\begin{example}\label{exagen} $A=1$ could be a blood pressure lowering medicine, $M$
blood pressure, and $Y$ the occurrence of a heart attack. To
investigate whether $A=1$ also has a direct effect on heart attacks,
one could do mediation analysis. Suppose that $M_0=\alpha^{(0)}_0+\alpha_1 C+e_0$, and $M_1=\alpha^{(1)}_0+\alpha_1 C+e_1$, and suppose that $e_0\sim e_1$, $e_0$ and $e_1$ are random error terms in $\mathbb{R}$ independent of $C$, and $\alpha_0^{(0)},\alpha_0^{(1)},\alpha_1\in\mathbb{R}$. Thus, treatment $A=1$ shifts the distribution of the blood pressure by $\alpha_0^{(1)}-\alpha_0^{(0)}$ without changing its shape. Suppose an intervention $I$ leads to $M_{1,I=1}=\alpha_0^{(2)}+\alpha_1C+e_{1,I=1}$. Then, $I$ satisfies equation~(\ref{defint}) if 1.\ $\alpha_0^{(2)}=\alpha_0^{(0)}$ (that is, $I=1$ shifts the distribution of the blood pressure, in the treated, by $\alpha_0^{(0)}-\alpha_0^{(1)}$ without changing its shape) and 2.\ $e_{1,I=1}\sim e_0$ is independent of $C$. Then, $M_{1,I=1}\sim M_0$, leading to (\ref{defint}). Intervention $I$ could for example be salt in a (possibly random) dosage depending on $C$. The effect of salt on heart attacks is believed to
be through its effect on blood pressure (see for example the CDC
website, http://www.cdc.gov/vitalsigns/Sodium/index.html), making
equation~(\ref{ident}) and thus Definition~\ref{mim} plausible
for this intervention. For natural direct and indirect effects, one would need to be able to shift the distribution of $M$ by $\alpha_0^{(0)}-\alpha_0^{(1)}$ without changing its shape, but additionally set $e_{1,I=1}=e_0$, resulting in $M_{1,I=1}=M_0$. For the direct and indirect effects introduced in \cite{Vanessa2}, one would randomize $e_1^I\sim e_0$ independent of $C$, and then need to set the mediator to $M_{1,I=1}=\alpha^{(0)}_0+\alpha_1C+e_{1,I=1}$.  \cite{Vanessa2} avoid the use of counterfactuals altogether using graphical models. Of the three interventions above, obviously, $e_1^I=e_0$ places the strongest restriction. It is related to the assumption of rank preservation sometimes made in the causal inference literature. Rank preservation also implies that two units with the same observed data have the same counterfactual data.

In this example, one could replace the fully parametric models by $M_0=g(C,e_0)$ and $M_1=g(C,e_1)+\beta$, with $g$ some function of $C$ and elements in $\mathbb{R}$. Then, $I$ needs to shift the distribution of $M_1$ given $C$ by $-\beta$. Or, one could have $M_0=g(C,e_0)$ and $M_1=g(C,e_1)+\beta_0+\beta_1C$, where now $I$ needs to shift the distribution of $M_1$ given $C$ by $-\beta_0-\beta_1C$.
\end{example}
If a pre-treatment common cause $\tilde{C}$ of mediator and outcome
has not been observed, equation~(\ref{ident}) without
$\tilde{C}$ is unlikely to hold. The reason is that the predictive
value of the mediator having a specific value under intervention $I$
is not the same as the predictive value of the mediator having a
specific value without intervention. The mediator $M_1$ under
treatment is predicted by the common cause $\tilde{C}$. However, if
$\tilde{C}$ is not included, under intervention $I$ the mediator
$M_{1,I=1}$ is not necessarily predicted by $\tilde{C}$.  So, the mediator
$M_1$ carries information on the common cause $\tilde{C}$, but the
mediator under intervention $I$, $M_{1,I=1}$, may not. Even if $M_{1,I=1}$
carries information on $\tilde{C}$, then the information on $\tilde{C}$
from $M_{1,I=1}=m$ may be different than the information on $\tilde{C}$
from $M_1=m$, because $M_{1,I=1}$ and $M_1$ have a different
distribution. As a consequence, the prognosis under treatment of units
with $M_1=m$ is different from the prognosis under treatment of units
with $M_{1,I=1}=m$, violating equation~(\ref{ident}).
Web-appendix~B has a detailed example of the
consequences of ignoring a pre-treatment common cause $\tilde{C}$ of
mediator and outcome.

When there is a post-treatment common cause $C'$ of the mediator and
the outcome, equation~(\ref{ident}) is also unlikely to hold. Assuming
that the intervention $I$ does not affect $C'$, the reason is the same
as for unobserved pre-treatment common causes. If the intervention $I$
also changes $C'$, basing mediation analysis on $I$ results in
estimating the effect mediated by $(C',M)$.

If an intervention $I$ satisfying equation~(\ref{defint}) is feasible,
which can be tested, equation~(\ref{ident}) or its relaxation could be
tested as well, by comparing the distributions of $Y_{1,I=1}$ given
$(M_{1,I=1},C)$ to the distribution of $Y_1$ given $(M_1,C)$. In order to
test this, an experiment must be carried out with three arms: ``do not
treat'', ``treat'', and ``treat under intervention $I$''. This is in
contrast with the existing literature on natural direct and indirect
effects, the assumptions of which can never be tested because they
involve the joint distribution of counterfactuals under different
treatments, which can never be jointly observed.

Now the organic direct and indirect effect of a treatment on the outcome can be defined:
\begin{defn}\emph{(Organic direct and indirect effect)}.\label{organic} Consider an organic intervention $I$. The organic direct effect of a treatment $A$ based on $I$ is
$E(Y_{1,I=1}) -E(Y_0)$. The organic indirect effect of a treatment $A$ based on $I$ is $E(Y_1)-E(Y_{1,I=1})$.
\end{defn}
Because the treatment is the same for both
$Y_1$ and $Y_{1,I=1}$, $E(Y_1)-E(Y_{1,I=1})$ is the organic
indirect effect, or mediated part of the effect. It is the effect of the organic intervention on the mediator, $I$, under $A=1$. If the distribution of the mediator does not depend on $A$, $I$ could be "no intervention on $M$", and the organic indirect effect is $0$. The organic indirect effect is also $0$ if the intervention on the mediator does not affect the outcome. Because the mediator has the same distribution for both
$Y_{1,I=1}$ and $Y_0$, $E(Y_{1,I=1}) -E(Y_0)$ is the organic
direct effect.
The direct effect is the effect of
treatment combined with an organic intervention as compared to no
treatment.
Very loosely, the direct effect is the effect of a
treatment that 1.\ has the same direct effect as treatment $A=1$: the
dependence of $Y_{1,I=1}$ on the covariates $C$ and on the mediator is the
same as that of $Y_1$, but 2.\ has no indirect effect through the
mediator (see (\ref{defint})). 
Notice that
$E(Y_1)-E(Y_0)=\left(E(Y_1)-E(Y_{1,I=1})\right)+\left(E(Y_{1,I=1})-E(Y_0)\right)$.
Thus, like for natural direct and indirect effects, organic direct and
indirect effects add up to the total effect of a treatment.
Organic direct and indirect effects often generalize natural direct and indirect effects:
\begin{thm} \label{gen}Under equation~(\ref{crosseqn}), natural direct and indirect effects and the direct and indirect effects defined in \cite{Vanessa2} are special cases of organic direct and indirect effects.
\end{thm}

\section{Uniqueness of organic direct and indirect effects}\label{unique}

Definition~\ref{organic} of organic direct and indirect effects
depends on the organic intervention $I$ and on the choice of
baseline common causes of mediator and outcome $C$. Although the
definitions of natural direct and indirect effects also depend on the
intervention (the mediator is set to a specific value), this has not
usually been made explicit. I argued that $C$ has to include all common causes of mediator and
outcome for equation~(\ref{ident}) to be plausible, and thus for an
intervention $I$ to be organic. This section formalizes the notion of
common causes of mediator and outcome, and argues that the organic
direct and indirect effects do not depend on (a) for given $C$, the
choice of organic intervention $I$ or (b) on the choice of common causes
$C$ of mediator and outcome, even if more than one set of common
causes exists.

Define a common cause of mediator and outcome given $C$ as follows:
\begin{defn}\label{common}(\emph{common cause}). $X$ is \emph{not} a common cause of mediator and outcome given $C$ if either equation~(\ref{common1}) or equation~(\ref{common2}) holds:
\begin{equation}\label{common1}
X\cip M_0\mid C \hspace{2cm} {\rm and}\hspace{2cm} X\cip M_1\mid C
\end{equation}
\begin{equation} \label{common2}
X\cip Y_1\mid M_1,C.
\end{equation}
\end{defn}
\noindent That is, $X$ is \emph{not} a common cause if, given $C$, either $X$
does not predict the mediator, or, given the mediator, $X$ does not
predict the outcome. In graphical language: $X$ is \emph{not} a common cause of
outcome and mediator if in a DAG that has $C$, $X$, $M$, and $Y$,
there either is no arrow from $X$ to $M$, or there is no direct arrow from
$X$ to $Y$. This definition is in line with, for example,
\cite{Pea}. If (all given $C$) $X$ predicts the mediator and, given the
mediator, $X$ predicts the outcome, it is a common cause of mediator
and outcome, and usually needs to be included in $C$ for
equation~(\ref{ident}) to hold with $C$ (see the discussion below
Definition~\ref{mim}). The following theorem is proved in Web-appendix~A:
\begin{thm} \label{same} For given $C$, the organic direct and indirect effect do not depend on the choice of organic intervention $I$ with respect to $C$. Furthermore, if $C$ and $\tilde{C}$ are different sets of common causes of mediator and outcome, $C$ is not a common cause of mediator and outcome given $\tilde{C}$, and $\tilde{C}$ is not a common cause of mediator and outcome given $C$, then the organic direct and indirect effect do not depend on whether the intervention is organic with respect to $C$ or organic with respect to $\tilde{C}$.
\end{thm}
Thus, if we restrict ourselves to interventions that are organic with respect to ``complete'' common causes $C$ (given $C$, any other pre-treatment covariate $X$ is not a common cause), organic direct and indirect effects are unique, and one can speak of ``the'' organic direct and indirect effect.

\section{Identifiability and estimation of organic direct and indirect effects}
\label{estimation}

When the treatment is randomized, $E(Y_1)$ and $E(Y_0)$ can simply be
estimated by the averages of $Y_1$ and $Y_0$ among units receiving
treatment and not receiving treatment, respectively. Therefore, in
order to estimate the organic direct and indirect effects of a
randomized treatment, this section focuses on estimating the
expectation of $Y_{1,I=1}$.
The following theorem is the main result of this article:
\begin{thm}\label{effects}\emph{(Organic direct and indirect effects: the mediation formula for randomized experiments).} Under randomized treatment and Definition of organic interventions~\ref{mim}, the following holds for an intervention $I$ that is organic with respect to $C$:
\begin{equation*}
E\left(Y_{1,I=1}\right)=\int_{(c,m)}E\left[Y\mid M=m,C=c,A=1\right]f_{M\mid C=c,A=0}(m)f_C(c)dm\,dc.
\end{equation*}
\end{thm}
\noindent Notice that to estimate $E\left(Y_{1,I=1}\right)$, only the distribution of $M$ under $A=0$ and of
$Y$ under $A=1$ are needed. Thus, Theorem~\ref{effects} can be used
both in the absence and in the presence of treatment-mediator
interaction (where the expectation of $Y$ depends on $M$
differently with or without treatment).
Theorem~\ref{effects} provides the same mediation formula as the previous
literature (see Section~\ref{natural}).
This formula depends on observable quantities only, and can be
estimated using standard models. The contribution of the current
article is to show that the definition and thus the
interpretation of direct and indirect effects, as well as the
conditions under which estimators for these effects are meaningful, can
be considerably relaxed.



\section{Estimating organic direct and indirect effects in observational studies}
\label{observational}

So-far, treatment was randomized.
This section extends the identification to
non-randomized treatments $A$. As before, $A$ is treatment, which is $1$
for the treated and $0$ for the untreated.
I adopt the usual consistency assumption (see e.g.\ \cite{Aids})
relating the observed to the counterfactual data:
\begin{assu} \emph{(Consistency).} \label{cons} If $A=1$, $M=M_1$ and $Y=Y_1$. If $A=0$, $M=M_0$ and $Y=Y_0$.
\end{assu}
For observational data, I allow that there exist baseline
covariates $Z$ (beyond the common causes of mediator and outcome, $C$)
that need to be included in the analysis in order to eliminate
confounding:
\begin{assu}\emph{(No Unmeasured Confounding).}\label{randobs}
\begin{equation*}A\cip \left(Y_1,M_1\right)\mid C,Z \;\;\;\;\;{\rm and}\;\;\;\;\; A\cip Y_0\mid C,Z\;\;\;\;\;{\rm and}\;\;\;\;\;
A\cip M_0\mid C,Z.
\end{equation*}
\end{assu}
\noindent Thus, given the measured pre-treatment
covariates $C$ and $Z$, treatment should not depend on the prognosis
of the units with or without treatment. For Assumption~\ref{randobs}
to hold, it is sufficient that $(C,Z)$ includes all the common causes
of the treatment, the mediator, and the outcome. This is a particular
representation of the usual assumption of no unmeasured confounding in
causal inference (see
e.g.~\citeauthor{Aids},\citeyear{Aids}). Assumption~\ref{randobs}
cannot be tested statistically. Subject matter experts have to
indicate whether they believe enough pre-treatment unit
characteristics have been observed in order for
Assumption~\ref{randobs} to be plausible.

Under Assumption~\ref{randobs}, the expectation of $Y_1$ and $Y_0$ can
be estimated using marginal structural models, the G-computation formula, or structural nested models.
Thus, I focus on the expectation of $Y_{1,I=1}$.
Section~\ref{definition} argued that in order for an intervention
to be organic with respect to $C$, $C$ usually has to include all common
causes of outcome and mediator. Therefore, if an extra $Z$ was
necessary for Assumption~\ref{randobs} of no unmeasured confounding to
hold, I will assume that given $C$, $Z$ is not a common cause of
mediator and outcome, as defined in
Definition~\ref{common}. Then,
\begin{thm}\label{effectsobs}\emph{(Organic direct and indirect effects: the mediation formula for observational studies).} Assume No Unmeasured Confounding Assumption~\ref{randobs}, Consistency Assumption~\ref{cons}, intervention $I$ is organic with respect to $C$ as in Definition~\ref{mim}, and given $C$, $Z$ is not a common cause of mediator and outcome as in Definition~\ref{common}. Then
\begin{eqnarray*}\lefteqn{E\left(Y_{1,I=1}\right)=}\\
&&\int_{(c,z,m)}E\left[Y\mid M=m,C=c,Z=z,A=1\right]f_{M\mid C=c,Z=z,A=0}(m)f_{C,Z}(c,z)dm\,d(c,z).
\end{eqnarray*}
\end{thm}
\noindent The proof is in Web-appendix~A. The resulting organic direct and indirect effects are similar to Theorem~\ref{effects}, in terms of observable quantities only, and can be estimated using standard methods.


%
%
%
%

\section{Application: Mother-to-child transmission of HIV/AIDS}\label{mtct}

HIV-infection can be transmitted from an HIV-positive mother to her
infant in utero, during birth, and by breast feeding. The rate of
HIV-transmission can be lowered by avoiding breast feeding, as well as
by treatments such as antiretroviral treatment (ART) and zidovudine
(AZT). ART and AZT lower the amount of HIV-virus, the HIV viral load,
in the blood of the mother. \cite{Sperling} describe that the effect
of AZT on mother-to-child transmission of HIV-infection is
surprisingly large, given the limited effect of AZT on the HIV viral
load in the blood of the mother. They estimated that less
than $20\%$ of the effect of AZT on mother-to-child transmission is due to the effect of AZT on the mother's HIV viral load, but
their analysis was not based on current notions of direct and indirect
effects.





This section describes how one could investigate how much of the
effect of AZT on mother-to-child transmission is mediated by the
effect of AZT on the HIV viral load in the blood of the mother (from now
on, the HIV viral load). I argue that the organic direct and indirect
effects defined in this article are well-defined and identified in
this situation, whereas natural direct and indirect effect are
undefined.

Suppose one would like to investigate the likely effect on
mother-to-child transmission of a potential new treatment that has the
same effect on HIV viral load as AZT but no direct effect on the
child's HIV status. Potentially, a low dosage of some type of ART could be such treatment.
Let $I$ be an intervention that, without AZT
treatment, causes the distribution of HIV viral load to
be the same as under AZT treatment; $I$ represents the potential new
treatment. Here, in contrast to most of the literature on mediation analysis, which focuses on the effect of an intervention on the mediator under treatment, interest focuses on the effect of an intervention on the mediator under \emph{no} treatment. In order to directly apply the method described in this article, we therefore re-code $A=0$ if a person was treated with AZT, and $A=1$ if a person was not treated with AZT. In the case of a linear model without treatment-mediator interaction,
$E[Y\mid M=m,A=a,C=c]=\beta_0+\beta_1m+\beta_2a+\beta_3^\T c$ (no term $\beta_4am$),
both approaches lead to the same direct effect, $\beta_2$, and
therefore also to the same indirect effect. In general, both
approaches can lead to different results. \cite{Tyler} and Web-appendix~C
discuss when each definition is most useful; this depends on the
context of the investigation.

In this example, one would expect that if AZT has a direct effect on
mother-to-child transmission, a mother's adherence to AZT treatment
is a post-treatment common cause of both HIV viral load $M$ and
mother-to-child transmission $Y$, because both $M$ and $Y$ will be
reduced under better adherence. Thus, one seems to need the
post-treatment covariate ``adherence'', $ad$, in $C$. However,
equation~(\ref{ident}) seems reasonable without compliance: if all
pre-treatment common causes are in $C$, so adherence is not a proxy
for other confounders, $ad\cip (Y_1,M_1)|C$ (recall $1$ indicates no treatment in this section). If adherence is not an issue for $I$, $ad\cip (Y_{1,I=1},M_{1,I=1}|C$. And if it is: because $I$ does not have a direct effect on
mother-to-child transmission, $Y_{1,I=1}\cip ad|M_{1,I=1},C$. Thus, if equation~(\ref{ident}) holds
with $ad$ in the conditioning event and all pre-treatment common
causes are in $C$, equation~(\ref{ident}) will also hold without
$ad$.

For ease of exposition, suppose that AZT treatment is randomized (the
approach can be generalized to observational studies as in
Section~\ref{observational}). I now illustrate how to use the identification result of
Section~\ref{estimation} to estimate the
indirect effect of AZT on mother-to-child transmission. Suppose that
$M_1\sim M_0+\beta_1+\beta_3^\T C\mid C$ holds for $M$ equal to log HIV
viral load.  Suppose in addition that the probability of
mother-to-child transmission without treatment follows a logistic
regression model of the form ${\rm
  logit}(Y=1\mid M=m,C=c,A=1)=\theta_0+\theta_1^\T M+\theta_2^\T C$.  Notice that
one only needs such a model for mother-to-child transmission under $A=1$ (no
treatment in this case). Then, by Web-appendix~D, it follows that
\begin{equation}E\bigl(Y_{1,I=1}\bigr)
=E\left[1/(1+{\rm exp}(-\theta_0-\theta_1(M-\beta_1-\beta_3^\T C)-\theta_2^\T C))\mid A=1\right].\label{expmtct}
\end{equation}
This expression can be estimated as indicated in
Web-appendix~D.  This leads to an estimator for the indirect
effect that does not use data on the outcomes for treated
mothers.

In contrast to the organic direct and indirect effects, the natural
direct and indirect effects are undefined in this application. They
involve $Y_{1,M_0}$, whether or not a newborn is infected without AZT
but with the HIV viral load of the mother set to the value it would
have had under AZT ($A=0$ here). How one could set the mediator to the value under
AZT is unclear. One can imagine treatments, for example low-dose ART, that have the same effect
on HIV viral load as AZT, as needed for organic direct and indirect
effects. However, it is unlikely that such a treatment would, for all
mothers, set the HIV viral load to the exact same value it would have
had under AZT. If AZT were a combination of substances, some
combination of a substance that affects HIV viral load and another
substance that might directly affect mother-to-child transmission, one
could imagine setting the HIV viral load to involve only the substance
that affected HIV viral load. However, like many treatments, AZT is
just one substance. I therefore conclude that for a treatment like
AZT, the organic direct and indirect effects are more natural than
their natural counterparts.

\section{Discussion}
\label{Discussion}

This article shows that, in contrast to the assumptions behind natural
direct and indirect effects, cross-worlds quantities and setting the mediator are not necessary
to define causal direct and indirect effects. This leads to newly
defined organic direct and indirect effects. Furthermore, this article
proves that, in contrast to natural direct and indirect effects,
identification of organic direct and indirect effects does not rely on
the existence of counterfactual outcomes under all combinations of the
treatment and the mediator. For identifiability of organic direct and
indirect effects, a distributional assumption linking the distribution
of the outcome under an organic intervention to the data replaces the
cross-worlds assumption which identifies natural direct and indirect
effects. This article focuses on organic interventions $I$, which
cause the distribution of the mediator given $C$ to be the same as
$M_0$, rather than setting the mediator value to $M_0$, as in natural
direct and indirect effects. In applications in the health or social
sciences, like epidemiology or psychology, one often wants to consider
which part of the effect of a treatment is mediated through some
covariate or trait. For example, one may want to investigate how much
of the effect of antiretroviral treatment, ART, on AIDS-defining
events and death is mediated by the CD4 count. In this example, it is
easier to envision an intervention that causes the CD4 count to have a
particular distribution rather than setting the CD4 count to a
specific value for each patient. If interventions on the mediator are
inconceivable, both natural and organic direct and indirect effects
are undefined.


I have shown that the proposed organic direct and indirect effects are
identified by the same expressions as developed previously in the
literature for natural direct and indirect effects. The contribution
of this article is to show that these mediation formulas hold in
substantially more generality. As a consequence, estimators based on
the mediation formulas have a much broader causal interpretation than
previously shown. The new definitions introduced in this article are
easy to interpret and can therefore be easily discussed with subject
matter experts. For an intervention $I$ to be organic it has to be
that, given pre-treatment characteristics $C$, the outcome under
treatment ``for a unit with $M_1=m$ under treatment'' is
representative of the outcome under treatment ``if the organic
intervention $I$ caused $M_{1,I=1}=m$''. This can be interpreted as that,
under treatment, the organic intervention has no direct effect on the
outcome.

An organic intervention $I$ is a considerable relaxation of $M_{1,I=1}=
M_0$. Still, it may be difficult to find an organic
intervention. Notice, however, that if there is an intervention
$\tilde{I}$ such that equation~(\ref{ident}) holds, then in some cases
it may be possible to construct an organic intervention $I$ by
adapting the dosage of $\tilde{I}$ as a function of $C$
(deterministically or randomly) in a way such that equation~(\ref{defint})
holds. If there is interest in figuring out what might be the benefit
of an intervention with only a direct or only an indirect effect, if
such an intervention would be developed in a lab, organic direct and
indirect effects are of interest. In any case, being able to actually
carry out organic interventions is not necessary to identify and
estimate organic direct and indirect effects.  Rather, organic
interventions can be employed as thought experiments useful to frame
the analysis and define the parameters of interest.



Natural direct and indirect effects are defined at the individual
level as well as the population level, whereas organic direct and
indirect effects are defined only at the population level. This
reflects that an organic intervention does not set the mediator to a
pre-specified value for each unit. 


In related work, the appendix of \cite{Tyler2012} considers
interventions that cause the mediator to have the same distribution as
without treatment, conditional on $C$. However, for identification
he still assumes existence of all $Y_{a,m}$ and
$Y_{a,m}\cip M\mid A,C,L$,
where $L$ is a post-treatment common cause of mediator and outcome.
This is problematic if one cannot set the mediator to particular
values. 



Following the previous literature, this article studies interventions that
do not affect pre-treatment common causes of mediators and outcomes. For example, inherited risk factors are thought to be
common causes of low birth weight and infant mortality. For
equation~(\ref{ident}) to be plausible, such common causes of mediators and outcomes
have to be taken into account. The consequences of
ignoring common causes are illustrated in Web-appendix~B for
the direct effect of smoking on infant mortality, not mediated by low
birth weight. The importance of observing common causes was also noted in e.g.~\cite{Imai}.


\cite{JamieThomas} argue that the natural direct effect, which they
call pure direct effect, ``is non-manipulable relative to $A$, $M$ and
$Y$ in the sense that, in the absence of assumptions, the pure direct
effect does not correspond to a contrast between treatment regimes of
any randomized experiment performed via interventions on $A$, $M$ and
$Y$.'' Organic direct and indirect effects are not subject to that
caveat. If there exists an organic intervention $I$ (not necessarily
$M_{1,I=1}=M_0$), then the organic direct and indirect effect induced by
$I$ are identified from the experiments ``do not treat'', ``treat'',
and ``treat under intervention $I$.'' Both conditions for $I$ to be
organic can be tested on the basis of these experiments, and the
organic direct and indirect effects do not depend on the choice of
organic intervention $I$.

Under an agnostic model, which does not assume the existence of
counterfactual outcomes, the natural direct and indirect effects are
obviously not defined: they are based on the cross-worlds
counterfactuals $Y_{1,M_0}$. In contrast, if an organic intervention
$I$ exists, the organic direct and indirect effects could have been
equivalently defined without counterfactual outcomes, because they can
be defined on the basis of interventions; see Web-appendix~F for details.

In future work, I will show that in contrast to natural direct and indirect effects, organic direct and indirect effects can be extended to provide an identification result for the case where there are
post-treatment mediator-outcome confounders. This will provide another
alternative to the three quantities described in \cite{Tylmed14}.

The methodology in this article could also be applied to the
study of the effects of future treatments based on prior data.
Consider, for example, the effect of a future treatment to lower
immune activation ($M$ in the notation of this article) in
HIV-positive patients. Suppose that this future treatment is aimed to
eventually prevent clinical events ($Y$). Assume that under the future
treatment, patients would have a specific distribution of immune
activation $M_1$, and the future treatment has no direct effect on the outcome $Y$:
conditional on a set of covariates $C$, the prognosis of patients
under the future treatment, $Y_1$, is the same as the prognosis $Y$ of
patients with the same immune activation in the observed data (compare with (\ref{ident})).  Then
the mean outcome under the future treatment can be estimated using a sample
counterpart of
\begin{equation*}E(Y_1)=\int_{(c,m)}E\left[Y\mid M=m,C=c\right]
f_{M_1\mid C=c}(m)f_{C}(c)dm\, dc.
\end{equation*}
The proof follows the same lines
as Theorem~\ref{effects}, but the interpretation is different because
the future treatment does not necessarily cause immune activation to
have the same distribution as some existing treatment $A$. Of course,
since the identifying assumption of equation~(\ref{ident}) cannot be
verified without experimental data of the future treatment, an
experiment with the future treatment would be necessary to confirm
this result. This last formula also provides
a mathematical underpinning of the application in \cite{Naimi}, who estimate the controlled direct
effect of an intervention when "only a portion of the population's mediator is altered".

To conclude, this article introduces organic direct and indirect
effects and provides identification and estimators for
these effects. The assumptions are weaker than for natural direct and
indirect effects.


\section{Funding}

This work was supported by the National Institutes of Health [grant number R01 AI100762]. The content is solely the responsibility of the author and does not necessarily represent the official views of the National Institutes of Health.

\section*{Acknowledgements}

The author thanks Eric Tchetgen Tchetgen and Tyler VanderWeele for
sharing their drafts and comments, Victor DeGruttola for comments
and for suggesting the mother-to-child HIV transmission example, and Alberto Abadie for extensive
comments on this article.

\bibliographystyle{biorefs}
\bibliography{ref}

\pagebreak

\begin{figure}[htb!]
\begin{picture}(390,100)
\thicklines
\put(180,100){\makebox(0,0){Figure~1: DAG summarizing the data.}}
\put(45,55){\makebox(0,0){$A$}}
\put(185,55){\makebox(0,0){$M$}}
\put(325,55){\makebox(0,0){$Y$}}
\put(255,5){\makebox(0,0){$C$}}
\thicklines
\put(65,53){\vector(1,0){100}}
\put(205,53){\vector(1,0){100}}
\put(255,12){\vector(-3,2){50}}
\put(255,12){\vector(3,2){50}}
\qbezier(55,60)(185,100)(310,60)
\put(310,60){\vector(2,-1){3}}
\end{picture}
\end{figure}

\noindent Because treatment $A$ is randomized, the pre-treatment covariate $C$
is not a cause of $A$ in the DAG.

\pagebreak

\appendix

\section{Web-appendix A: Proofs}\label{Proofs}

\noindent{\bf Proof of Theorem~4.3:} (4.3) is trivial for $M_{1,I=1}=M_0$. (4.3) also follows immediately if $M_{1,I=1}$ is a random draw of $M_0$ given $C$. Furthermore, it is easy to
see that for $M_{1,I=1}=M_0$, and if all $Y_{a,m}$ are well-defined, then
cross-worlds Assumption~(3.1) implies
equation~(4.4): for $M_{1,I=1}=M_0$, equation~(4.4) states
$Y_{1,m}\mid M_0=m,C=c \sim Y_{1,m}\mid M_1=m,C=c$, and under
equation~(3.1) for randomized treatment $A$, $Y_{1,m}$
depends, for given $C$, neither on $M_0$ nor on $M_1$. For $M_{1,I=1}$ a random draw, equation~(4.4) states $Y_{1,m}\mid M_{1,I=1}=m,C=c \sim Y_{1,m}\mid M_1=m,C=c$, and under
equation~(3.1) for randomized treatment $A$, $Y_{1,m}$
depends, for given $C$, not on any mediators.

\noindent {\bf Proof of Theorem~5.2:}
First, let $I$ be an intervention that is organic with respect to $C$. Then
\begin{eqnarray}\label{org}
E\left(Y_{1,I=1}\right)
&=&E\left(E\left[Y_{1,I=1}\mid M_{1,I=1},C\right]\right)\nonumber\\
&=&\int_{(c,m)}E\left[Y_{1,I=1}\mid M_{1,I=1}=m,C=c\right]f_{M_{1,I=1}\mid C=c}(m)dm\,f_C(c)dc\nonumber\\
&=&\int_{(c,m)}E\left[Y_1\mid M_1=m,C=c\right]f_{M_0\mid C=c}(m)dm\,f_C(c)dc.
\end{eqnarray}
In this proof, the first two equalities follow from the definition of
conditional expectation. The third equality follows from
Definition~4.1, (4.3) and (4.4). Thus, the
choice of $I$ does not influence the direct and indirect effect, as
long as it is organic with respect to $C$.

Next, let $I^C$ be an intervention that is organic with respect to $C$ and $I^{\tilde{C}}$ and
intervention that is organic with respect to $\tilde{C}$. I assumed that $C$ is not a common cause
of mediator and outcome given $\tilde{C}$, and $\tilde{C}$ is not a common cause of mediator and outcome given $C$;
hence there are 4 different cases, with either (5.5) or (5.6) holding for $C$ and $\tilde{C}$, respectively.
I will show that under any of the 4 different cases,
\begin{equation*}
E\left(Y_1^{I^{\tilde{C}}}\right)=\int_{(\tilde{c},m,c)}E\left[Y_1\mid M_1=m,\tilde{C}=\tilde{c},C=c\right]f_{M_0\mid \tilde{C}=\tilde{c},C=c}(m)f_{\tilde{C},C}(\tilde{c},c)dc\,dm\,d\tilde{c}.
\end{equation*}
Since the conditions and the result are symmetric in $C$, $\tilde{C}$, it follows that also
\begin{equation*}
E\left(Y_1^{I^{C}}\right)=\int_{(\tilde{c},m,c)}E\left[Y_1\mid M_1=m,\tilde{C}=\tilde{c},C=c\right]f_{M_0\mid \tilde{C}=\tilde{c},C=c}(m)f_{\tilde{C},C}(\tilde{c},c)dc\,dm\,d\tilde{c}.
\end{equation*}
But then, $E\left(Y_1^{I^{\tilde{C}}}\right)=E\left(Y_1^{I^{C}}\right)$.

Suppose first that $C\cip M_0\mid \tilde{C}$ and $C\cip M_1\mid \tilde{C}$. Then
\begin{eqnarray*}
\lefteqn{E\left(Y_1^{I^{\tilde{C}}}\right)}\\
&=&\int_{(\tilde{c},m)}E\left[Y_1\mid M_1=m,\tilde{C}=\tilde{c}\right]f_{M_0\mid \tilde{C}=\tilde{c}}(m)f_{\tilde{C}}(\tilde{c})dm\,d\tilde{c}\\
&=&\int_{(\tilde{c},m,c)}E\left[Y_1\mid M_1=m,\tilde{C}=\tilde{c},C=c\right]f_{C\mid M_1=m,\tilde{C}=\tilde{c}}(c)dc\,f_{M_0\mid \tilde{C}=\tilde{c}}(m)f_{\tilde{C}}(\tilde{c})dm\,d\tilde{c}\\
&=&\int_{(\tilde{c},m,c)}E\left[Y_1\mid M_1=m,\tilde{C}=\tilde{c},C=c\right]f_{C\mid \tilde{C}=\tilde{c}}(c)f_{M_0\mid \tilde{C}=\tilde{c},C=c}(m)f_{\tilde{C}}(\tilde{c})dc\,dm\,d\tilde{c}\\
&=&\int_{(\tilde{c},m,c)}E\left[Y_1\mid M_1=m,\tilde{C}=\tilde{c},C=c\right]f_{M_0\mid \tilde{C}=\tilde{c},C=c}(m)f_{\tilde{C},C}(\tilde{c},c)dc\,dm\,d\tilde{c}.
\end{eqnarray*}
The first line follows from equation~(\ref{org}). The second line conditions on $C$.
In the third line I changed the order of integration and used $C\cip M_0\mid \tilde{C}$ and $C\cip M_1\mid \tilde{C}$.


Alternatively, suppose that $C\cip Y_1|M_1,\tilde{C}$. Then,
\begin{eqnarray*}
\lefteqn{E\left(Y_1^{I^{\tilde{C}}}\right)}\\
&=&\int_{(\tilde{c},m)}E\left[Y_1\mid M_1=m,\tilde{C}=\tilde{c}\right]f_{M_0\mid \tilde{C}=\tilde{c}}(m)f_{\tilde{C}}(\tilde{c})dm\,d\tilde{c}\\
&=&\int_{(\tilde{c},m)}E\left[Y_1\mid M_1=m,\tilde{C}=\tilde{c}\right]\int_c f_{M_0\mid \tilde{C}=\tilde{c},C=c}(m)f_{C|\tilde{C}=\tilde{c}}(c)dc\,f_{\tilde{C}}(\tilde{c})dm\,d\tilde{c}\\
&=&\int_{(\tilde{c},m),c}E\left[Y_1\mid M_1=m,\tilde{C}=\tilde{c}\right]f_{M_0\mid \tilde{C}=\tilde{c},C=c}(m)f_{C\mid \tilde{C}=\tilde{c}}(c)f_{\tilde{C},c}(\tilde{c},c)dc\,dm\,d\tilde{c}\\
&=&\int_{(\tilde{c},m),c}E\left[Y_1\mid M_1=m,\tilde{C}=\tilde{c},C=c\right]f_{M_0\mid \tilde{C}=\tilde{c},C=c}(m)f_{\tilde{C},c}(\tilde{c},c)dc\,dm\,d\tilde{c}.
\end{eqnarray*}
The first line follows from equation~(\ref{org}). The second line conditions on $C$.
The last line follows from $C\cip Y_1\mid M_1,\tilde{C}$.
\hfill $\Box$\\

\noindent {\bf Proof of Theorem~6.1:}
\begin{eqnarray*}
E\left(Y_{1,I=1}\right)
&=&\int_{(c,m)}E\left[Y_1\mid M_1=m,C=c\right]f_{M_0\mid C=c}(m)dm\,f_C(c)dc\\
&=&\int_{(c,m)}E\left[Y_1\mid M_1=m,C=c,A=1\right]f_{M_0\mid C=c,A=0}(m)dm\,f_C(c)dc\\
&=&\int_{(c,m)}E\left[Y\mid M=m,C=c,A=1\right]f_{M\mid C=c,A=0}(m)f_C(c)dm\,dc.
\end{eqnarray*}
The first equality follows from equation~(\ref{org}). The second
equality follows from the fact that treatment was randomized; this
implies that
\begin{equation*}A\cip \left(Y_{1},M_1\right)\mid C \hspace{1cm}{\rm and}\hspace{1cm}A\cip M_0\mid C.
\end{equation*}
The last equality follows from the randomization.
\hfill $\Box$\\

\noindent {\bf Proof of Theorem~7.3:} Theorem~7.3 assumed that either equation~(5.5) or equation~(5.6) holds for $Z$.
Suppose first that equation~(5.5) holds for $Z$. Then
\begin{eqnarray*}
\lefteqn{E\left(Y_{1,I=1}\right)}\\
&=&\int_{(c,m)}E\left[Y_1\mid M_1=m,C=c\right]f_{M_0|C=c}(m)dm\,f_C(c)dc\\
&=&\int_{(c,m)}\int_z E\left[Y_1\mid M_1=m,Z=z,C=c\right]f_{Z\mid M_1=m,C=c}(z)dz\,f_{M_0\mid C=c}(m)f_C(c)dm\,dc\\
&=&\int_{(c,z,m)} E\left[Y_1\mid M_1=m,Z=z,C=c,A=1\right]f_{Z\mid C=c}(z)f_{M_0\mid Z=z,C=c}(m)f_C(c)dz\,dm\,dc\\
&=&\int_{(c,z,m)} E\left[Y_1\mid M_1=m,Z=z,C=c,A=1\right]f_{M_0\mid Z=z,C=c,A=0}(m)f_{C,Z}(c,z)dz\,dm\,dc\\
&=&\int_{(c,z,m)} E\left[Y\mid M=m,Z=z,C=c,A=1\right]f_{M\mid Z=z,C=c,A=0}(m)f_{C,Z}(c,z)dm\,dz\,dc.
\end{eqnarray*}
The first line follows from equation~(\ref{org}). The second line conditions on $Z$.
The third line uses (5.5), for both $M_0$ and $M_1$, Assumption~7.2, and changes the order of integration.
The fourth line follows from Assumption~7.2. The last line follows from Assumption~7.1.

Next, suppose that equation~(5.6) holds for $Z$.
\begin{eqnarray*}
\lefteqn{E\left(Y_{1,I=1}\right)}\\
&=&\int_{(c,m)}E\left[Y_1\mid M_1=m,C=c\right]f_{M_0\mid C=c}(m)dm\,f_{C}(c)dc\\
&=&\int_{(c,m)}E\left[Y_1\mid M_1=m,C=c\right]\int_z f_{M_0\mid Z=z,C=c}(m)f_{Z\mid C=c}(z)dz\,dm\,f_C(c)dc\\\
&=&\int_{(c,z,m)}E\left[Y_1\mid M_1=m,Z=z,C=c\right]f_{M_0\mid Z=z,C=c}(m)f_{Z,C}(z,c)dm\,dz\,dc\\
&=&\int_{(c,z,m)}E\left[Y_1\mid M_1=m,Z=z,C=c,A=1\right]f_{M_0\mid Z=z,C=c,A=0}(m)f_{Z,C}(z,c)dm\,dz\,dc\\
&=&\int_{(c,z,m)}E\left[Y\mid M=m,Z=z,C=c,A=1\right]f_{M\mid Z=z,C=c,A=0}(m)f_{Z,C}(z,c)dm\,dz\,dc.
\end{eqnarray*}
The first line follows from equation~(\ref{org}). The second line follows by conditioning on $Z$.
The third line follows from (5.6) and changing the order of integration.
The fourth line follows from Assumption~7.2. The fifth line follows from Assumption~7.1.
\hfill $\Box$\\

\section{Web-appendix B: The smoking-and-low-birth-weight paradox}\label{smokingapp}

Section~4 argued that if a common cause of mediator and
outcome $\tilde{C}$ has not been observed, it is often not reasonable to
think that equation~(4.4) without $\tilde{C}$ would hold.
As an example, this appendix considers the case of maternal smoking and
infant mortality. The effect of smoking during pregnancy ($A=1$) on
infant mortality may be mediated by low birth weight. It turns out
that a naive analysis leads to the conclusion that the direct effect
of maternal smoking on infant mortality is beneficial. \cite{SonMig}
explain this ``birth weight paradox'' and provide an explanation for
the possible biases. This appendix shows how this relates to the setup
of this article.

For exposition simplicity assume that whether a pregnant woman
smokes or not is unrelated to her prognosis with respect to low birth
weight or complications in her infant in the ``smoking'' and ``not
smoking'' scenarios. So, differences in outcomes between smokers and
nonsmokers are caused by smoking only, effectively implying that the
treatment ``smoking'' can be considered randomized. In practice, this
may be violated if women with other unhealthy behaviors besides
smoking are more likely to smoke. Those complications are ignored
here, because the issues addressed in this appendix are present even
under randomized treatment, and relaxing the randomization assumption
was already discussed in Section~7.

Some infants may have a low birth weight due to genetically determined
birth defects, which are likely not caused by smoking, or due to
environmental causes other than smoking like malnutrition. These
causes may be more predictive of infant mortality than smoking
(\cite{SonMig}). For exposition simplicity this appendix bases the
discussion on genetically determined birth defects as common causes of
birth weight and infant mortality. Denote these by
$\tilde{C}$. Suppose that, as in most studies, $\tilde{C}$ is not
observed. Now consider an intervention $I$ (Definition~4.1
equation~(4.3)) which causes birth weight for the smoking
mothers to have the same distribution as the birth weight for
non-smoking mothers, without changing genetically determined birth
defects $\tilde{C}$. Then, the prognosis of an infant had the mother
smoked and ``had the infant had a normal birth weight $M_{1,I=1}$ under
intervention $I$'' is most likely not the same as the prognosis of an
infant had the mother smoked and ``had the infant had normal birth
weight $M_1$ without intervention''. Without the intervention, in an
infant of a smoking mother with normal birth weight $M_1$, genes
responsible for birth defects are most likely more favorable: the
birth weight was normal without intervention, even while the mother
was smoking. So, one would think that the prognosis $Y_1$ is good for such
an infant. Under intervention $I$, some of the infants of smoking
women with normal birth weight $M_{1,I=1}$ will have genetically
determined birth defects: the birth weight has been intervened on to
be normal without changing genetically determined birth defects. The
possibility of genetically determined birth defects would lead to a
worse prognosis $Y_{1,I=1}$ for such infants. Thus, equation~(4.4)
will generally not hold in this situation.


Next, I consider how this issue affects the estimators of the direct
and indirect effects if $\tilde{C}$ is ignored (which it has to be,
because it is assumed that $\tilde{C}$ is unobserved). Let the outcome
$Y$ be an indicator of infant mortality, and let $I$ be an
intervention for which equation~(4.3) holds. If $\tilde{C}$
is ignored, $E(Y_{1,I=1})$ would be estimated using the data for women who
smoked but who had infants with relatively high birth weights, because
that is the distribution of the birth weights $M_{1,I=1}$ under
intervention $I$.  As argued in the previous paragraph, this approach
is too optimistic, and thus the mortality probability $E(Y_{1,I=1})$ is
underestimated. Thus, the part of the effect of smoking that is
mediated through low birth weight, the indirect effect of smoking, is
overestimated. As a consequence, the direct effect of smoking on
infant mortality is underestimated.

This is in line with what was found in e.g.\ \cite{SonMig}, who
studied controlled direct effects, and found that conditional on birth
weight, smoking and infant mortality were negatively associated in
infants with low birth weight. A naive approach would thus conclude
that the direct effect of smoking is beneficial. \cite{SonMig}
explained this by noting that low birth weight may be more harmful if
caused by genetic birth defects than if caused by smoking. As outlined
above, this is a violation of equation~(4.4).

The solution to this issue is to try to include in $C$ as many
pre-treatment common causes of mediator and outcome as feasible. In
the case of the genetically determined birth defects in the above
example, this could perhaps be done through observed traits of the
newborn babies. If this is unfeasible, conclusions may be flawed
because equation~(4.4) fails to hold. The direction of the
bias can be reasoned as described in the previous paragraph: in this
example, ignoring birth defects results in an overestimation of the
organic indirect effect of smoking (mediated by birth weight), and an
underestimation of the detrimental organic direct effect of smoking
(not mediated by birth weight).

The discussion in this section illustrates the importance of the
assumptions behind mediation analysis. One can compare whether the
distribution on the left hand side of equation~(4.4) puts more
mass on larger values of the outcome or on smaller values of the
outcome as compared to the distribution on the right hand side. Thus,
an advantage of the current approach is that the direction of the bias
that results from lack of validity of equation~(4.4) can be
discussed in the context of each particular application.

\section{Web-appendix~C: Interventions on the mediator under treatment or on the mediator under no treatment?}\label{0or1}

There has been some discussion in the previous literature about
whether one should consider setting the mediator to its value under
treatment versus setting it to its value without treatment (see
e.g.~\citeauthor{Tyler}, \citeyear{Tyler}). As indicated in
Section~8, the approach in this article can easily be
extended to incorporate both. To illustrate what might be of most
clinical interest in a particular setting, consider two scenarios. In
scenario 1, an alternative treatment $I'$ changes the mediator the
same way conventional treatment $A$ does, without having a direct
effect on the outcome. This would be especially relevant for example
if the direct effect of treatment $A$ is a harmful side effect. In
this case, one would want to compare the distribution of the outcome
under no treatment with the distribution of the outcome under no
treatment if the mediator under intervention $I'$, $M_{0,I'=1}$, has the same
distribution as $M_1$. For example, with $Y_{0,I'=1}$ the outcome under
$I'$ with $A=0$, one would want to estimate
$E\bigl(Y_{0,I'=1}-Y_0\bigr)$ as the effect mediated through $M$. This
is a different quantity than the organic indirect effect
of Section~4, but can be estimated in a similar way by changing the coding of $A$
as in Section~8. In scenario 2, the quantity of interest is
the effect of an alternative treatment $\tilde{A}$, where $\tilde{A}$ has
the same direct effect as treatment $A$, but does not affect the
mediator $M$. This would be especially relevant if the effect of
treatment $A$ on the mediator is a harmful side effect. In this case
one would want to consider an intervention such that $M_{1,I=1}$ under
treatment has the same distribution as $M_0$. In that situation, the
quantity of interest is $E\bigl(Y_{1,I=1}-Y_0\bigr)$, the organic direct
effect of treatment $A=1$ as defined in
Definition~4.1. Scenario~1 motivates an intervention that
causes the mediator without treatment to have the same distribution as
$M_1$, scenario~2 motivates an intervention that causes the mediator with
treatment to have the same distribution as $M_0$. When studying the
biological mechanisms by which particular treatments are effective,
both types of interventions may be of interest.

\section{Web-appendix~D: Inference under randomized treatment}
\label{inference}

I now illustrate how one might use the identification result of
Section~6 to estimate $E\left(Y_{1,I=1}\right)$, and hence
the organic indirect and direct effects, under semi-parametric
assumptions.


Suppose that $M_1\sim M_0+\beta_1+\beta^\T_3C\mid C$, with
$\beta_1\in\mathbb{R}$ and $\beta_3\in\mathbb{R}^k$. This would
be the case if, as in e.g.~\cite{Valeri}, $M$ follows a regression model
$M=\beta_0+\beta_1A+\beta_2^\T C+\beta_3^\T AC+\epsilon$, where the random
variable $\epsilon$ has the same distribution given $C$ under
treatment as without treatment, and with $\beta_1\in\mathbb{R}$ and
$\beta_2,\beta_3\in\mathbb{R}^k$. Suppose in addition that the
expected value of $Y$ given $C$ and $M$ under treatment follows some
parametric model of the form
$E\left[Y\mid M=m,C=c,A=1\right]=f_\theta(m,c)$.  Notice that this last
model applies only to the distribution of $Y$ conditional on $A=1$,
not conditional on $A=0$.  This implies that the model does not
restrict treatment-mediator interactions. Then, Theorem~6.1 implies
$E\left(Y_{1,I=1}\right)$ \linebreak
$=E\left[f_\theta(M-\beta_1-\beta_3^\T C,C)\mid A=1\right]$
(proof: see below). This can be estimated by fitting the
models for $\beta$ and $\theta$ using standard methods, plugging the
parameter estimates in, and replacing the expectation given $A=1$ by
its empirical average. Standard errors can be estimated with the
bootstrap.

Notice that the resulting estimator uses changes in the distribution
of the mediator with and without treatment, but the distribution of
the outcome only in treated units. This leads to an estimator for the
indirect effect that does not use data on the outcomes for untreated
units.

\cite{Valeri} provide code to estimate direct and indirect effects based on the mediation formula for the case where $M$ and $Y$ both follow regression or logistic regression models.

\noindent {\bf Proof of inference under randomized treatment:}
\begin{eqnarray*}
E\left(Y_{1,I=1}\right)
&=&\int_{(c,m)}E\left[Y\mid M=m,C=c,A=1\right]f_{M\mid C=c,A=0}(m)f_C(c)dmdc\nonumber\\
&=&\int_{(c,m)}f_\theta(m,c)f_{M\mid C=c,A=1}(m+\beta_1+\beta_3^\T c)f_C(c)dmdc\nonumber\\
&=&\int_{(c,\tilde{m})}f_\theta(\tilde{m}-\beta_1-\beta_3^\T c,c)f_{M\mid C=c,A=1}(\tilde{m})f_C(c)d\tilde{m}dc\nonumber\\
&=&E\left[f_\theta(M-\beta_1-\beta_3^\T C,C)\mid A=1\right],
\end{eqnarray*}
where the first equality follows from Theorem~6.1, the
second equality follows from $M_1\sim M_0+\beta_1+\beta_3^\T C\mid C$,
see above, the third equality from a change of variables with $\tilde{m}=m+\beta_1+\beta_3^\T
c$, and the fourth equality from the fact that treatment $A$ is
randomized, and therefore the distribution of $C$ does not depend on
$A$. \hfill $\Box$\\

\section{Web-appendix~E: Organic direct and indirect effects: independence assumptions instead of distributional assumptions}

Some readers may be more at ease with independence assumptions underlying causal inference than with the
distributional assumptions considered in the main text. This can be done in the current context as follows.
Let $R$ describe the possible treatments as follows: $R=0$: treatment $0$, $R=1$: treatment $1$
and $R=2$: treatment $1$ combined with an "organic" intervention $I$ on the mediator. Equivalent to the definition
in the main text, the definition for $I$ being an organic intervention on the mediator could be formulated as
that both equations~(\ref{defint3}) and~(\ref{ident3}) are satisfied:
\begin{equation}\label{defint3}
M\cip R\mid C=c,R\neq 1
\end{equation}
\begin{equation}\label{ident3}
Y\cip R\mid M=m, C=c, R\neq 0.
\end{equation}
Of course, for easier interpretation, $R\neq 1$ could be replaced by "$R=0$ or $R=2$" and $R\neq 0$
could be replaced by "$R=1$ or $R=2$". The first of these assumptions states that, for given pre-treatment
covariates $C$, the mediator is independent of whether the mediator was intervened on during treatment
versus no treatment was given. The second of these assumptions states that, for given mediator and pre-treatment
covariates $C$, the outcome is independent of whether the mediator got its value $m$ because it was intervened on
during treatment versus treatment $1$ was given.

\section{Web-appendix~F: Organic direct and indirect effects without counterfactuals}

Some of the literature on causal inference is avoiding counterfactuals, see e.g.~\cite{Dawid}, \cite{Vanessa2}, and~\cite{Geneletti}. Although this has not been a concern in the main manuscript, some readers may appreciate that organic direct and indirect effects can also be defined without counterfactuals, if "organic" interventions are possible in a three-arm clinical trial with $R=0$: treatment $0$, $R=1$: treatment $1$ and $R=2$: treatment $1$ combined with an "organic" intervention $I$ on the mediator. In this setting, the definition for $I$ being an organic intervention on the mediator is that both equations~(\ref{defint2}) and~(\ref{ident2}) are satisfied:
\begin{equation}\label{defint2}
M\mid R=2,C=c \sim M \mid R=0,C=c
\end{equation}
\begin{equation}\label{ident2}
Y\mid R=2,M=m,C=c \sim Y\mid R=1,M=m,C=c.
\end{equation}
Equation~(\ref{defint2}) states that the distribution of the mediator under treatment combined with the intervention $I$ is as under treatment $0$,
and equation~(\ref{ident2}) intuitively states that the intervention $I$ on the mediator has no direct effect on the outcome $Y$.

The organic direct and indirect effects based on $I$ can now be defined as
\begin{equation*}
E[Y|R=1]-E[Y|R=2]
\end{equation*}
and
\begin{equation*}
E[Y|R=2]-E[Y|R=0]
.\end{equation*}
As in the main paper, the mediation formula holds for $E[Y\mid R=2]$ because
\begin{eqnarray*}
E[Y\mid R=2]
&=&E(E[Y|M,C,R=2])\\
&=&\int_{m,c}E[Y|M=m,C=c,R=2]f_{M|C=c,R=2}(m)f_{C|R=2}(c)\\
&=&\int_{m,c}E[Y|M=m,C=c,R=1]f_{M|C=c,R=0}(m)f_{C}(c),
\end{eqnarray*}
because $R$ is randomized, (\ref{defint2}), and~(\ref{ident2}).

\end{document}